**Title**

**Enhanced electron injection for efficient proton acceleration and neutron production in femtosecond laser-driven nano-structured targets**

**Authors**

Yingzi Dai[1,2]†, Chengyu Qin[1]†, Hui Zhang[1]*, Guoqiang Zhang[3], Changbo Fu[4], Xiangai Deng[4], Dirui Xu[1], Shuai Xu[1], Xuesong Geng[1], Jing Wang[1,2], Bowen Zhang[1], Yunwei Cui[1], Xiaojing Guo[1], Weifu Yin[3], Yanqi Liu[1], Xingyan Liu[1], Cheng Wang[1], Zongxin Zhang[1], Bingnan Shi[1], Lianghong Yu[1], Xiaoyan Liang[1], Yuxin Leng[1], Baifei Shen[5], Liangliang Ji[1]‡, Ruxin Li[1]

**Affiliations**

[1]State Key Laboratory of Ultra-intense laser Science and Technology, Shanghai Institute of Optics and Fine Mechanics, Chinese Academy of Sciences (CAS), Shanghai, China.
[2]Center of Materials Science and Optoelectronics Engineering, University of Chinese Academy of Sciences, Beijing, China.
[3]Shanghai Institute of Applied Physics, Chinese Academy of Sciences (CAS), Shanghai, China.
[4]Key Laboratory of Nuclear Physics and Ion-beam Application (MOE), Institute of Modern Physics, Fudan University, Shanghai, China.
[5]Department of Physics, Shanghai Normal University, Shanghai, China.
Corresponding author. Email: * zhanghui1989@siom.ac.cn; ‡ jill@siom.ac.cn;
†These authors contributed equally to this work.

**Abstract**

Micro- or nano-structured targets are advantageous in enhancing and manipulating laser-proton acceleration, due to the increased absorption of laser energy and onset of direct laser acceleration for high-energy electrons. Here, we experimentally demonstrate that nano-wire-array printed on a flat substrate is an efficient nano-injector of relativistic electrons that leads to a significant boost of laser-driven proton acceleration and neutron production beyond normal geometry. By employing an ultra-intense ($2\times10^{21}$ W/cm$^2$) femtosecond laser pulse to irradiate nano-wire-array targets, protons with cut-off energies of 62.8 MeV are generated, and notably, the energy conversion efficiency from laser to protons reaches up to 9% - 3.5 times higher than that of flat foils. After bombarding a beryllium converter, $1.1\times10^{10}$ neutrons are produced. Full 3D particle-in-cell simulations have reproduced experimental results and reveal interference mechanisms between the nano-wires and substrate, leading to continuous pumping of electrons from the substrate and standing-wave enhanced re-injection from the wire tip. This efficient injection finally results in the large sheath field and thus high yield of energetic protons and neutrons. Dependence on the wire length and scaling with laser amplitude are further discussed. These results suggest that 3D-printed structures are promising in developing compact laser-driven high-flux proton and neutron sources for numerous applications.

**Teaser**

Nano structures driven by lasers are demonstrated to efficiently generate high-flux ultra-short proton and neutron sources.

# MAIN TEXT

## Introduction

The generation of energetic ion beams using ultra-intense, ultra-short laser pulses has attracted considerable interests due to their potential applications in fast ignition fusion *(1)*, cancer therapy *(2)*, proton radiography *(3)*, nuclear physics etc. These applications are further enabled by femtosecond petawatt-class lasers which usually provide high laser intensity and repetition rate *(4,5)* compared to picosecond systems. In particular, the ultra-short nature of laser-driven particles is promising in developing ultra-fast neutron pulses (ps~ns duration) by impinging on catcher targets *(6-8)* - a novel source for compact neutron imaging and spectroscopy *(9,10)*. The "pitcher–catcher" scheme involves two stages: ion acceleration and nuclear reaction, in which light ions (protons or deuterons) are typically used as projectiles to bombard heavy targets (primarily lithium or beryllium) *(11,12)*. Therefore, enhancing the laser-proton conversion efficiency (CE), especially the number of protons at moderate energies, is crucial for developing high-yield laser-driven neutron sources.

Several mechanisms have been proposed for laser-driven proton acceleration, including target normal sheath acceleration (TNSA) *(13-15)*, collisionless shock acceleration (CSA) *(16,17)*, radiation pressure acceleration (RPA) *(18-22)*, magnetic vortex acceleration *(23,24)*, and those induced by relativistic transparency *(25,26)*. Recently, proton energies beyond 100 MeV have been achieved via hybrid mechanisms *(27,28)*. For applications of laser-driven proton sources, one usually employs TNSA since it is the most reliable and reproducible approach in current experiments. In TNSA, protons are accelerated by a quasi-static electric sheath field formed by hot electrons at the rear surface of solid targets. However, when using femtosecond laser pulses, the interaction is confined to the skin-depth region, resulting in limited laser energy absorption and inefficient ponderomotive electron temperatures *(29)*. This can be mitigated by tuning the pre-plasma condition of flat targets *(30,31)*. While the theoretical limit of laser-to-proton CE is predicted to be ~8% *(32)*, typical experimental results with femtosecond lasers lie in the range of 1~4% *(33,34)*, posing a significant barrier for high-impact applications such as proton-driven neutron generation.

To address this, structured targets, such as nanorod or 'velvet' *(35,36)*, nano sphere *(37)*, micro cylinder/tube *(38,39)*, foam layer *(40)*, near-critical-density (NCD) material *(41-43)*, and wire array *(44-49)*, have been explored to improve laser energy absorption *(50,51)*, electron generation and secondary radiation. These structures increase the effective interaction area and facilitate direct laser acceleration (DLA) of electrons. The laser pulse penetrates through the low-density or structured region and generates a large number of relativistic electrons with energies that exceed the ponderomotive scaling *(39,52,53)*, thereby strengthening the sheath field and enhancing proton acceleration. In particular, the nano/micro-arrayed structures further provide an approach to manipulate electron sources for various purposes. In these scenarios, the structured components are mostly fabricated on flat substrates. Thus, the sheath field is seeded by electron sources from both the structure and substrate. These contributions are typically regarded as additive, where possible interference between the array and the substrate is usually neglected, which could reshape injection, transport, and sheath formation for electron acceleration.

In this work, we identify and investigate an interference regime in which the interplay between the nano-structure and the substrate leads to enhanced electron injection for efficient ion acceleration. By using the ultra-intense femtosecond laser pulse to irradiate 3D-printed nano-wire array (NWA) structures mounted on flat foils *(54)*, proton beams with cut-off energies above 60 MeV and yields of $4.5\times10^{12}$ (> 1 MeV) are obtained at an optimal nano-wire height, corresponding to a CE of 9%. Compared with the case of flat foils, this marks a 3.5-fold improvement in CE and a two-fold increase in cut-off energy. When these beams are directed onto a beryllium converter, neutron production via the $^9$Be(p, xn) reaction reaches up to $1.1\times10^{10}$, over twice that from flat foils. Full 3D particle-in-cell (PIC) simulations confirm these findings and reveal that with an appropriate nano-wire height, substrate electrons are continuously drawn into the nano-wire structure and re-injected through the wire tip, forming a nano-scale injector of relativistic electrons. This injection is further enhanced by standing-wave fields established by substrate reflection. These synergistic effects lead to stronger sheath fields and more efficient ion acceleration. The effect of wire height on proton cut-off energy and acceleration efficiency is further discussed based on the injection geometry. Our results demonstrate a scalable and efficient scheme for generating high-energy proton beams and ultra-fast neutron pulses, opening a new avenue for applications in nuclear physics, high-energy-density science, and advanced radiography.

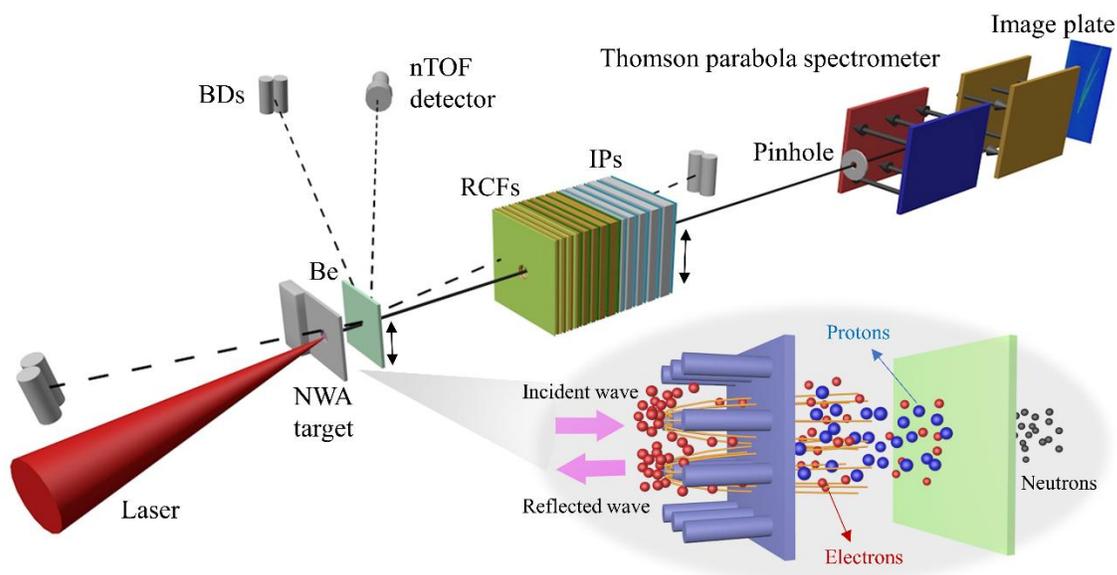

**Fig.1. Schematic of the experimental setup.** Protons are firstly accelerated by femtosecond laser interacting with NWA targets and then are deposited into a beryllium (Be) converter for neutron production. At the target normal direction, the radiochromic film (RCF) and image plate (IP) stacks are used to record the spatial profile of protons and electrons, respectively. Following that, a Thomson parabola (TP) spectrometer is employed to measure ion spectra. Meanwhile, the neutron Time-of-Flight (nTOF) detector and bubble detectors (BDs) are used to diagnose neutrons.

## Results
### Experimental Results
The schematic of the experimental setup is shown in Fig. 1. To diagnose accelerated protons, a stack of 5 × 5 cm HDv2 and EBT3 radiochromic films (RCFs) is placed 5.2 cm downstream from targets along the normal direction *(55)*. Behind RCF stacks, a BAS-SR image plate (IPs) stack is used to measure electron beam profiles *(56)*. Layers of copper and aluminum foils of designed thicknesses are inserted into RCF and IP stacks to attenuate proton and electron energies. The whole stacks are enwrapped with a 15-μm-thick aluminum foil for debris protection, and have a central 3-mm-diameter hole allowing direct proton transmission to a Thomson parabola (TP) spectrometer located 60 cm from target. To generate neutrons, a 1-cm-thick Be converter shielded by 15-μm-thick aluminum foil to eliminate radiation noises, is positioned 0.18 cm behind the target, which is mounted on a movable stage to separate proton and neutron signal measurement. The neutron Time-of-Flight (nTOF) detector, consisting of EJ309 liquid scintillator optically coupled to photomultiplier tubes, is used to measure the neutron spectrum and placed 14.4 m behind the nuclear reaction region at an angle of 64°. The signal of nTOF detector is recorded by a digital oscilloscope, where the signal of the laser driven x-ray serves as a time reference for the neutron energy analysis.

Fig. 2A shows proton spectra derived from the RCF stack. For the flat foil, the maximum proton cut-off energy is measured at 33 MeV. In contrast, the NWA target with a wire height of 2 μm achieves a significantly higher cut-off energy of 62.8 MeV — almost a two-fold increase. It can be clearly seen from Fig. 2B that all NWA targets with different heights exhibit better proton acceleration performance compared to the flat foil. The laser-to-proton CE (kinetic energy > 1 MeV) reaches the maximum value of 9% for the 3-μm-height NWA target. This represents a 3.5-fold enhancement over the flat foil case, and to our knowledge, sets a new benchmark for CE in femtosecond laser-driven proton acceleration. These energy and efficiency improvements are consistent across multiple shots. Here, the CE is obtained by integrating the proton spectra derived from the RCF stack shown in Fig. 2A and the absolute dose calibration is performed using a reference proton beam source to ensure quantitative accuracy *(57)*.

The electron spectra obtained from the IP stack for the flat foil and 2-μm-height NWA target are illustrated in Fig. 2C, showing that the NWA target produces ~$7.9 \times 10^{10}$ electrons (15 nC) for energy more than 7.5 MeV, nearly an order of magnitude higher than that of the flat foil (~$1.1 \times 10^{10}$, 2 nC). The electron temperature for the NWA case also shows a modest increase from 5.5 MeV to 6.4 MeV, indicating that the primary contribution to enhanced proton acceleration comes from the dramatic increase in relativistic electron population rather than in electron energy. It is worth noting that this observation is distinct from previous studies employing much longer nano-wire or foam targets, where electron energy/temperature enhancement is more significant *(40,58)*.

Fig. 2D presents neutron energy spectra from the nTOF detector for the flat foil and NWA target with the 2-μm height, respectively. For the 2-μm-height NWA case, the neutron yield reaches up to $1.1 \times 10^{10}$ per shot, which is more than doubled compared with the flat foil case ($4.9 \times 10^{9}$). Meanwhile, the neutron energy extends beyond 10 MeV, indicating that the high-energy proton component effectively induces (p, xn) reactions in the Be converter. Three bubble detectors (BDs) at different angles of 0°/90°/180° supplement the nTOF detector for yield measurements *(59)*, showing consistent enhancement using the NWA structure. More details about the measurement results can be seen in Materials and Methods.

To validate experimental results of neutron production, we simulate the neutron emission using Geant4 *(60)*, employing the experimentally measured proton spectra as input. The simulated neutron spectra show reasonable agreement with experimental results, both in spectral shape and absolute yield (seen in Fig. 2D). These results highlight not only the record-level CE achievable with optimal NWA targets, but also demonstrate their potential for generating compact, ultra-fast neutron sources. Such high-flux, short-duration neutron sources are valuable for applications in time-resolved neutron radiography, neutron spectroscopy, fusion diagnostics, and nuclear material interrogation.

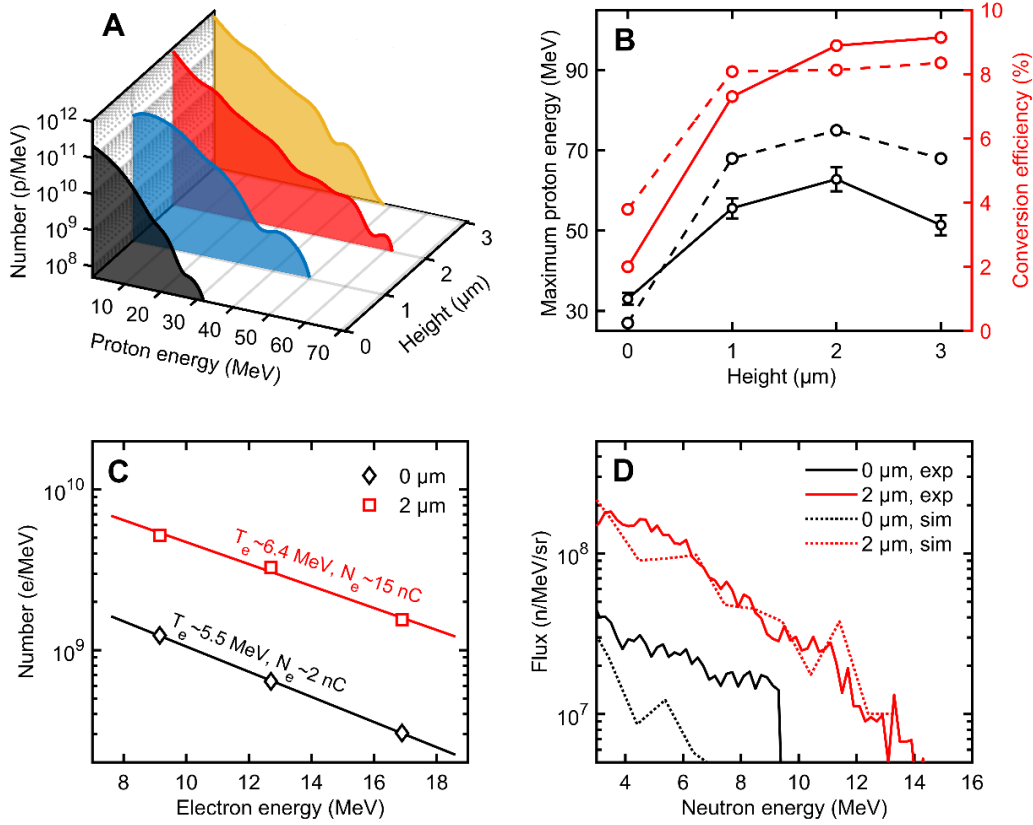

**Fig.2. Experimental results. (A)** Proton energy spectra obtained from RCF stacks for flat targets (0 μm) and NWA targets with three heights of 1, 2 and 3 μm, respectively. **(B)** Cut-off energy and conversion efficiency as a function of nano-wire height. The solid and dashed lines represent experimental and simulated results, respectively. The error bars are defined by the measurement deviation. The electron spectra obtained from the IP stack **(C)** and neutron energy spectra from the nTOF detector **(D)** for the flat foil and NWA target with the 2-μm height, respectively.

**Simulation Results**
The simulations reproduce the trend observed in experiments, indicating that NWA targets yield both higher proton cut-off energies and CE. Experimental results show that the CE basically saturates with longer wires while the cut-off energy peaks at h = 2 μm. To investigate the underlying mechanism, we simulated three distinct cases of bare substrate, bare wire array, and combined NWA target. Figs. 3A, B and C present the transverse electric field $E_y$ in the x-y plane at t = 69 fs, along with typical trajectories of electrons (> 10 MeV) initially located in nano-wires (Fig. 3B) and substrates (Figs. 3A and C). For the flat foil, electrons are injected and driven forward/backward primarily by the V×B force, or the ponderomotive acceleration *(61)*. This is restricted within the skin-depth region thus both the number and energy of injected electrons are limited, as shown in Fig. 3A. When the laser interacts with NWA structure, electrons are directly injected via the laser transverse electric field *(62)*, and then the laser magnetic field turns the transverse momentum to longitudinal momentum, forming an energetic beam through DLA *(63)*.

Fig. 3B indicates that such injection is most active at the tip of each wire, resulting in charge displacement and in turn leaving the nano-wire tip highly positively charged, thereby generating a significant potential gradient along the wire. Cold electrons located in the wire are dragged towards the tip, forming a strong return current inside the wire (will be shown later). The driven return current finally induces the azimuthal magnetic field, which can be employed to pinch the wire itself *(64)*.

Surprisingly, we find that cold electrons in the substrate foil also flow back to wires under the influence of the gradient of electric potential in wires, as clearly shown in Fig. 3C. The resulting electron reflux inside the nano-wire propagate to the wire-tip and inject into the wire interval. In other words, the composition of NWA and substrate forms an efficient nano-pump and injector for electrons. Without this mechanism, the injection would terminate since the accumulated charge-separation field eventually becomes strong enough to prevent electron escaping. Compared to that of a single 200-nm CH target, the number of hot electrons (> 10 MeV) significantly increases through the nano-injection mechanism. The relationship between the hot electron number and the simulation time is shown in Fig. 3D. It can be seen that ~$2.3\times10^{10}$ substrate electrons (black solid line) are accelerated beyond 10 MeV for the NWA case, which is nearly 2 times that of the flat foil (black dashed line).

The reflection of the incident laser from the substrate forms a standing-wave interference pattern in front of the target surface. As we can see from Fig. 3E that, the transverse electric field $E_y$ (normalized to the incident value $E_{y0}$) is about twice the original field strength in the NWA – substrate configuration. In contrast, for a single NWA without substrate, the peak amplitude remains nearly identical to the incident field. This enhanced transverse field further increases the electron injection at each wire tip. Fig. 3D shows that for both two cases, the number of hot electrons in the nano-wire array rises rapidly after t ≈ 50 fs—well above that in flat targets. Crucially, attaching a substrate significantly increases the hot electron yield from $6.8\times10^{10}$ to $1.5\times10^{11}$, indicating that substrate-induced field enhancement in the NWA target not only pulls more electrons from the wires but also enables them to gain greater energy from the laser pulse.

In NWA targets, hot electrons accelerated in target front propagate together with the laser pulse in a proper phase. When the laser reaches the substrate and is reflected, these hot electrons, which have gained considerable kinetic energy from laser, continue to propagate through the substrate, forming an enhanced sheath field at the rear side. In other words, this process is equivalent to the so-called enhanced TNSA process. The NWA target yields a total population of hot electrons of $1.9\times10^{11}$ including wire array and substrate contributions, representing an 8.6-fold enhancement over the bare substrate case, in line with experimental results shown in Fig. 2C.

To quantify the impact on proton acceleration, we introduced the electric field from a bare NWA without substrate into simulations of a bare substrate. Fig. 3F shows proton spectra for three configurations of bare substrate, bare substrate with an introduced electric field, and combined NWA target. The introduced electric field induces substrate electron reflux toward the target front, where subsequent energy gain occurs via DLA, which increases the maximum proton energy from 28 MeV in the bare-substrate case to 43 MeV. In the case of the combined NWA target, the maximum proton energy reaches 72 MeV. In other words, the synergistic effect between nanowire-induced electron re-flowing and substrate-introduced field enhancement ultimately leads to the optimized acceleration mechanism in the NWA target, thereby showing a trend similar to the increase observed in experiments (seen in Fig. 2B).

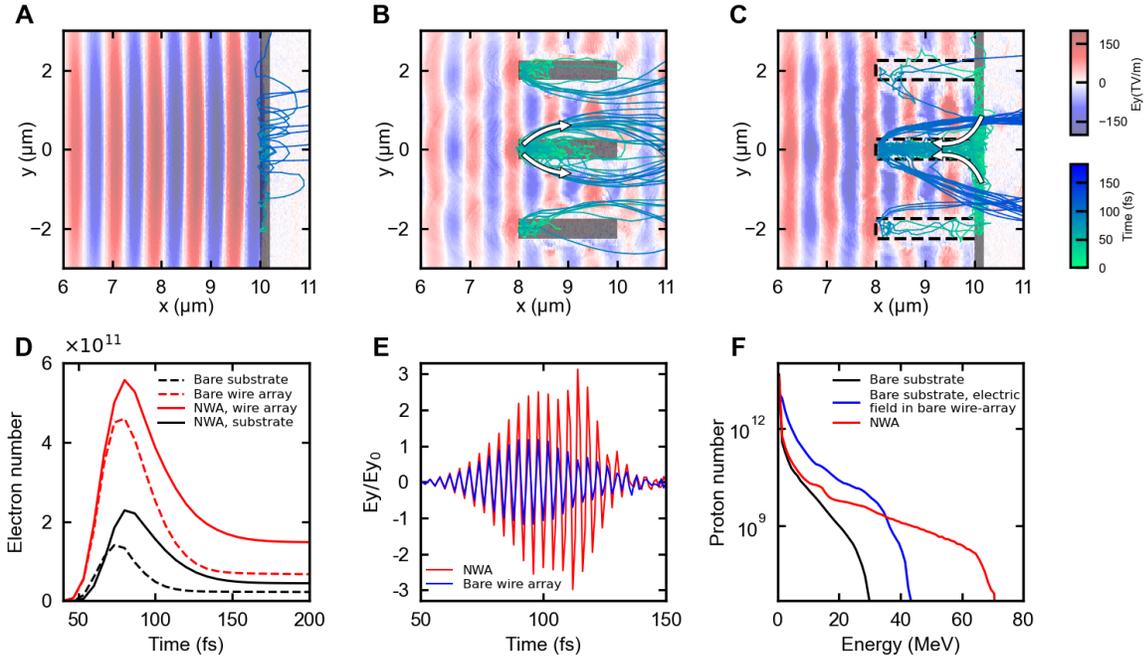

**Fig.3. Simulation results.** 2D slice of the transversal electric field $E_y$ at t = 69 fs and hot electron (> 10 MeV at t = 69 fs) trajectories in the x-y plane for three different cases of **(A)** bare substrate, **(B)** bare wire array, **(C)** combined NWA target. **(D)** Time evolution of the hot electron number with different target geometries. Solid lines (black: substrate, red: wire array) for the combined NWA target and dashed lines (black: bare substrate, red: bare wire array) for individual components. **(E)** Time evolution of the transversal electric field $E_y$ in the wire interval at x = 9 μm. **(F)** The proton spectra for three different cases of the bare substrate, the bare substrate with an introduced electric field, and the combined NWA target.

## Discussion

The previous analysis compared the flat foil and NWA target in terms of electron acceleration, intensifying the mechanism behind enhanced electron injection and the formation of strong longitudinal current. Figs. 4A-C show the spatial distribution of the longitudinal current $J_x$ for NWA targets of heights h =1 μm, 2 μm and 3 μm. The strong current due to reflow and injection of electrons in Fig. 4B is consistent with electron trajectories in Fig. 3, where electrons are injected from the wire tip at angles of $\theta_0 \approx 25°$ into the wire gap. In general, increasing wire height corresponds to more NWA mass and hence higher electron population, and also extends the acceleration path of extracted electrons, leading to higher electron energy. It is anticipated that the performance should be better when varying the height from 1 μm to 3 μm. This is true in the CE shown in Fig. 2B, where the highest value is obtained at 3 μm. The proton cut-off energy, however, shows a decrease when the wire height increases from 2 to 3 μm.

The reason becomes clear in Fig. 4D, which plots $J_x$ nearby the rear surface of the substrate at x = 10.5 μm and t = 69 fs. For the 2-μm-height wire, the peak value of $J_x$ exceeds 7 MA/μm$^2$, which drops to about 4 MA/μm$^2$ for 3 μm wires. This is ascribed to the focusing geometry of extracted electrons. For a wire pitch of p = 2 μm, the longitudinal distance for electrons from adjacent wires to converge is $L = \frac{(p-d)/2}{\tan(\theta_0)} = 1.6\ \mu m$. An insufficient wire height prevents proper convergence of adjacent electron streams, while an excessive wire height causes early converge, leading to electron beam divergence at the substrate surface. The 2-μm-height wire enables effective merge of these extracted electron bunches at the central region of the substrate front surface. We plot the near-field pattern of the longitudinal electric field ($E_x$) in Figs. 4 E-G. It is demonstrated that longer wires produce significantly broader acceleration field regions. However, compared to the 3-μm-height wire, the acceleration field of the 2-μm-height wire peaks higher in the focusing region but is slightly weaker elsewhere. Consequently, while the CE is slightly higher for the 3-μm-height case, the cut-off proton energy is maximized at h = 2 μm.

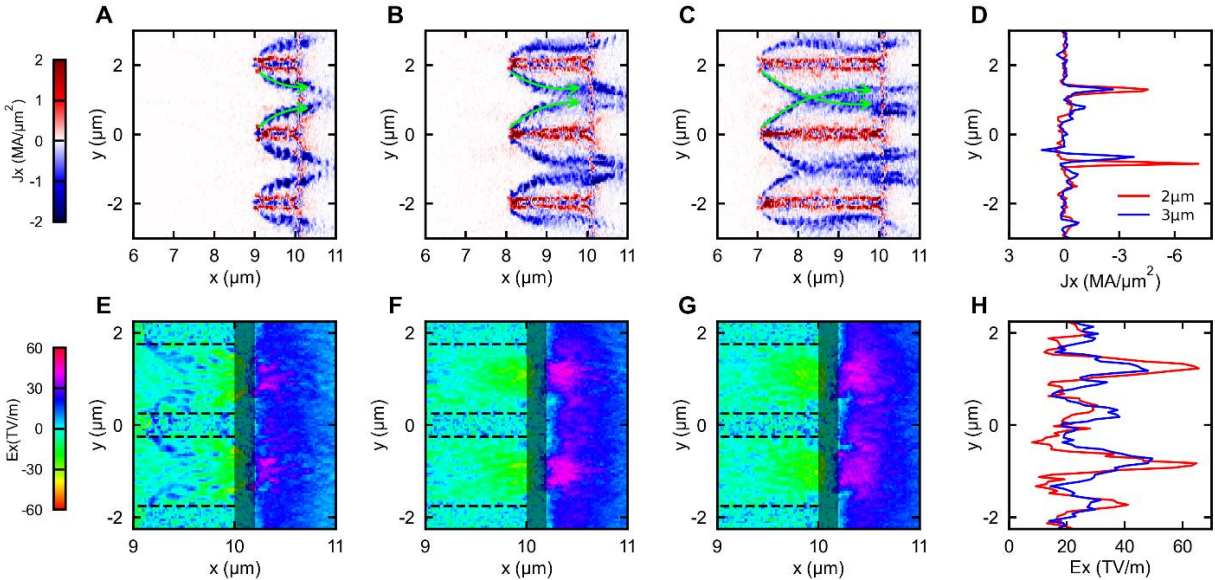

**Fig.4. The effect of the wire height on the longitudinal current and electric fields.** Spatial distribution of the longitudinal current $J_x$ in the x-y plane at t = 69 fs for three wire heights of **(A)** 1 μm, **(B)** 2 μm and **(C)** 3 μm. The green arrows in (A)-(C) represent the electron bunch direction. **(D)** Transversal profiles of $J_x$ at x = 10.5 μm. Spatial distribution of the longitudinal electric field $E_x$ in the x-y plane at t = 69 fs for three wire heights of **(E)** 1 μm, **(F)** 2 μm and **(G)** 3 μm. **(H)** Transversal profiles of $E_x$ at x = 10.5 μm.

In addition, we next examine how laser amplitude affects the CE of protons η and maximum proton energy $E_{max}$, as shown in Fig. 5. Varying the laser amplitude from 10 to 100, the CE is approximately proportional to the square root of the laser amplitude $\eta \propto a_0^{0.5}$, reaching 17.8% at $a_0 = 100$. The proton cut-off energy scales as $E_{max} \propto a_0^{1.5}$, higher than pure TNSA ($E_{max} \propto a_0$) but lower than RPA ($E_{max} \propto a_0^2$). Given the experimental challenge for RPA, the NWA structure offers a reliable approach for efficient generation of energetic protons and secondary particles. Based on the simulated proton spectra, the neutron yield is estimated using Geant4, which increases with the laser amplitude and eventually saturates at about $2.5\times10^8$ neutrons per joule. Here, the saturation indicates that the femtosecond petawatt laser is sufficient to produce the neutron source efficiently.

In conclusion, we have experimentally demonstrated that printing a NWA structure in front of a flat substrate via 3D nano-printing technique can markedly enhance both proton acceleration and neutron generation. In our experiment, the laser-to-proton energy CE reaches up to 9% with the cut-off energy beyond 60 MeV. The resulting neutron yield in $^9$Be(p, xn) reaction attains $1.1\times10^{10}$. 3D PIC simulations reveal that this enhancement originates from two coupled mechanisms: (i) continuous substrate–electron reflux into the nano-wires, acting as an efficient "nano pump–injector" that sustains hot-electron production, and (ii) standing-wave formation at the substrate front, which significantly boosts the transverse electric field and facilitates direct electron pull-out from the wire tip. We further find that varying the NWA wire height allows control over electron beam convergence, which directly influences the sheath-field strength and resulting proton energy. These results establish NWA–substrate geometry as compact, high-efficiency sources of energetic protons and high-flux neutrons, offering a scalable route toward laser-driven particle sources for applications in nuclear physics, materials science, and high-energy-density research.

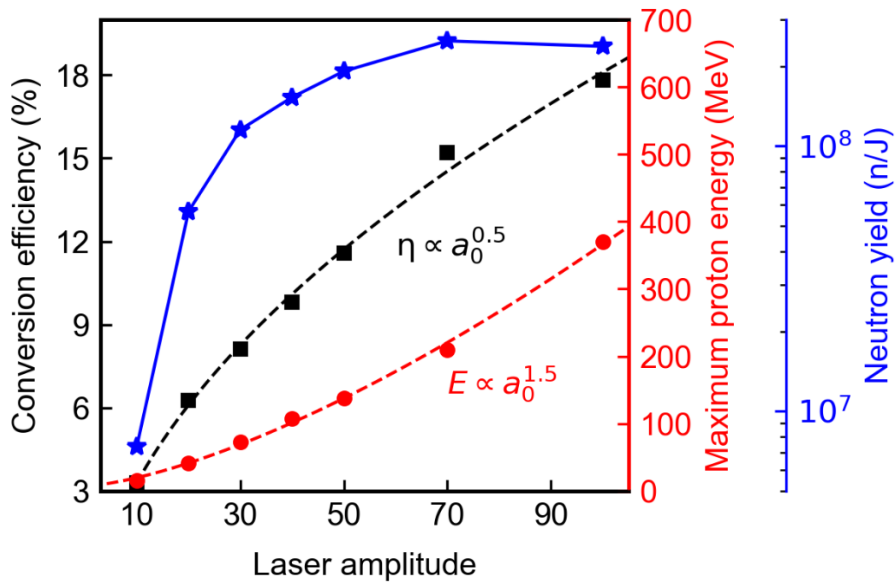

**Fig.5. Scaling with laser amplitude.** The proton conversion efficiency (black square), maximum proton energy (red circle) and neutron yield per joule (blue star) versus the input laser intensity for the 2-µm-height NWA target.

## Materials and Methods
### Laser system and NWA targets
The experiment was carried out in Shanghai Superintense Ultrafast Laser Facility *(65)*. A linearly polarized femtosecond laser pulse with energy of 55 J, pulse duration of 28 fs (FWHM, full width at half-maximum) and central wavelength of 800 nm was focused onto the target using an F/4 off-axis parabola (OAP) with an incidence angle of 10°. The focal spot is 5.8 µm in diameter (FWHM), resulting in a peak intensity of around $2\times10^{21}$

W/cm$^2$ corresponding to the normalized laser amplitude $a_0$ of 30 ($a_0 = eE/m_ec\omega$), where e and $m_e$ are the electron charge and mass, E is the laser electric field, c is the speed of light in vacuum and $\omega$ is the laser frequency, respectively. A single plasma mirror was inserted after OAP to enhance the laser contrast.

The NWA structure used in experiment is printed on 200-nm-thick CH flat foils using two-photon polymerization with a Nanoscribe 3D printer *(66)*. The polymer density is 1.17 g/cm$^3$. The diameter (d = 500 nm) and spatial period (p = 2 µm) of wire arrays are fixed while the height varies with three different cases (h = 1, 2, 3 µm).

**PIC simulations**

To elucidate the underlying physics, we carried out full 3D PIC simulations using the EPOCH code *(67)*. The simulation box is 50 µm × 22 µm × 22 µm in the x × y × z direction, with a cell size of 0.02 µm × 0.05 µm × 0.05 µm and 8 macro-particles per cell. The simulated parameters are chosen to match with experimental conditions. A linearly y-polarized Gaussian laser pulse with the normalized peak amplitude $a_0 = 30$, duration $\tau = 28$ fs, and spot size $\sigma = 5.8$ µm, propagates in the +x direction. The fully ionized CH foil is positioned at x = 10 µm with C:H = 1:3, thickness of 200 nm and electron density $216n_c$ ($n_c = m_e\varepsilon_0\omega_0^2/e^2$), where $n_c$ is the critical density, $\varepsilon_0$ and $\omega_0$ are the vacuum permittivity and the plasma frequency, respectively. The NWA attached to the front surface of the CH foil is modeled using the same composition and density. The preplasma distribution is introduced in both front and rear sides of the target as well as around the nano-wire structure with a scale length of 200 nm, extending the 2D model *(68)* to 3D.

**Measurement results from BDs**

Bubble detectors have emerged as often used diagnostic to supplement neutron measurement at laser-driven sources, which are insensitive to γ background and strong electromagnetic pulses, as their deposited energy is too low to trigger bubble formation. The response curve with neutron energy for BDs is dependent on the manufacturing process and can range from 10 keV to 10000 keV, with a reasonably flat dose response above 300 keV *(69)*.

Figure S1 shows the neutron yield data from three bubble detectors. For two cases of flat foil and 2-µm-height NWA target, the neutron yield shows similar angular distribution. Elevated neutron yields measured by the bubble detector at 90° (target side) may be attributed to the thicker converter target employed, which obstructs forward neutron emission. More importantly, we see that the average neutron influence has a high yield $2.41 \times 10^9$ n/sr in NWA target, which is more than two times that in flat foil, in good agreement with nTOF results.

Note that BDs have a detection energy threshold of 0.3 MeV, whereas the nTOF detector quantifies neutrons above 2.7 MeV, which is the reason that BDs indicate a neutron yield higher than that calculated by integrating energy spectra in Fig. 2(d). According to the simulated spectrum, the neutron number for > 2.7 MeV constitute ~ 36% of the population for > 0.3 MeV, unifying yield values obtained from both detection methods.

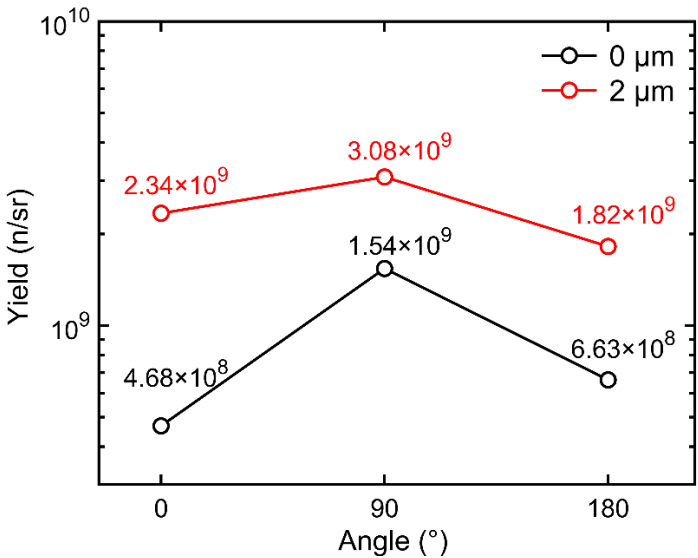

**Fig.6. Enhanced neutron yields characterized by BDs.** Neutron yields at three different angles of 0°, 90°, and 180° for two cases of the flat foil and the 2-μm-height NWA target.

**Acknowledgments**

The authors would like to express their sincere gratitude to the Shanghai Superintense Ultrafast Laser Facility for their professional technical support and experimental collaboration.

**Funding:** This work was supported by the Strategic Priority Research Program of the Chinese Academy of Sciences (Grant No. XDB0890300), the National Natural Science Foundation of China (Grants No. 12388102), the National Key R&D Program for Young Scientists (Grant No. 2024YFA1612700), the Shanghai Rising-Star Program (Grant No. 23QA1410600), the CAS Project for Young Scientists in Basic Research (Grant No. YSBR060), the Youth Innovation Promotion Association of Chinese Academy of Science (Grant No. 2021242) and the National Key R&D Program of China (Grant No. 2022YFE0204800).

**Author contributions:** Yingzi Dai, Chengyu Qin, Hui Zhang and Liangliang Ji conceptualized and wrote the manuscript, with feedback from all other authors. Hui Zhang, Baifei Shen, Liangliang Ji and Ruxin Li obtained funding and supervised this project. Hui Zhang, Chengyu Qin and Yingzi Dai performed the experiment, with assistance from Dirui Xu, Shuai Xu, Jing Wang, Bowen Zhang, Yunwei Cui, Xiaojing Guo and Weifu Yin. Yanqi Liu, Xingyan Liu, Cheng Wang, Zongxin Zhang and Bingnan Shi provided the technical support of laser facility. Experimental data analysis was conducted by Yingzi Dai, Chengyu Qin, Guoqiang Zhang, Changbo Fu and Xiangai Deng. Yingzi Dai performed and analyzed the PIC simulations, with assistance from Liangliang Ji and Xuesong Geng. Supervision was provided by Lianghong Yu, Xiaoyan Liang and Yuxin Leng.

**Competing interests:** Authors declare that they have no competing interests.

**Data and materials availability:** All data are available in the main text.

# Figures

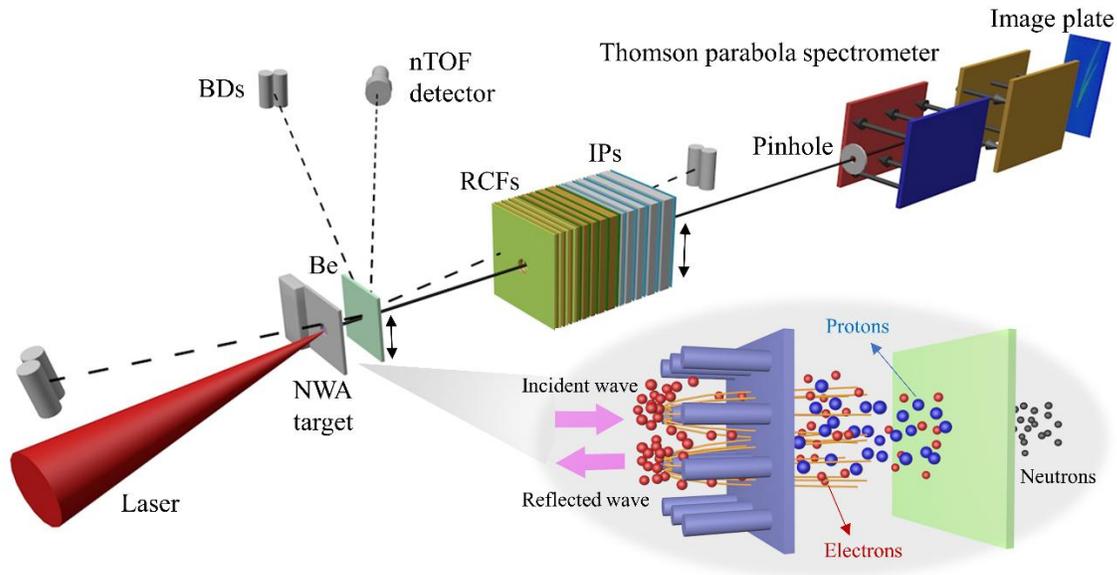

**Fig.1. Schematic of the experimental setup.** Protons are firstly accelerated by femtosecond laser interacting with NWA targets and then are deposited into a beryllium (Be) converter for neutron production. At the target normal direction, the radiochromic film (RCF) and image plate (IP) stacks are used to record the spatial profile of protons and electrons, respectively. Following that, a Thomson parabola (TP) spectrometer is employed to measure ion spectra. Meanwhile, the neutron Time-of-Flight (nTOF) detector and bubble detectors (BDs) are used to diagnose neutrons.

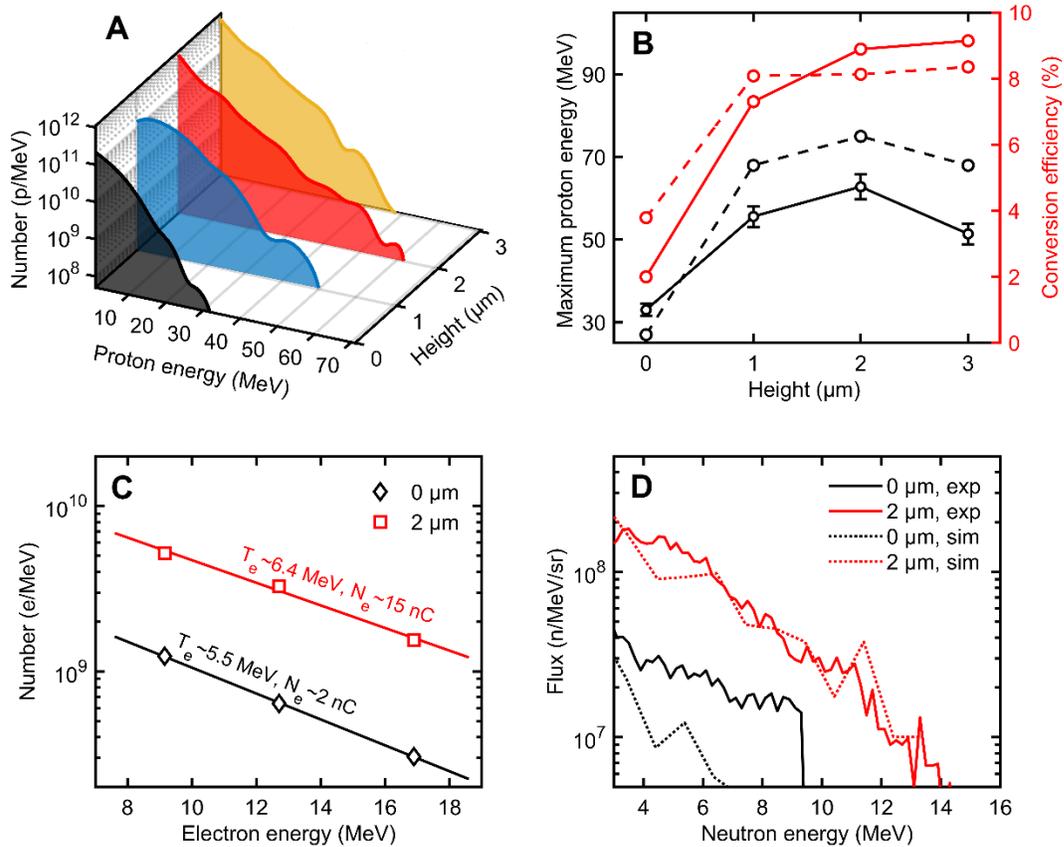

**Fig.2. Experimental results.** (**A**) Proton energy spectra obtained from RCF stacks for flat targets (0 μm) and NWA targets with three heights of 1, 2 and 3 μm, respectively. (**B**) Cut-off energy and conversion efficiency as a function of nano-wire height. The solid and dashed lines represent experimental and simulated results, respectively. The error bars are defined by the measurement deviation. The electron spectra obtained from the IP stack (**C**) and neutron energy spectra from the nTOF detector (**D**) for the flat foil and NWA target with the 2-μm height, respectively.

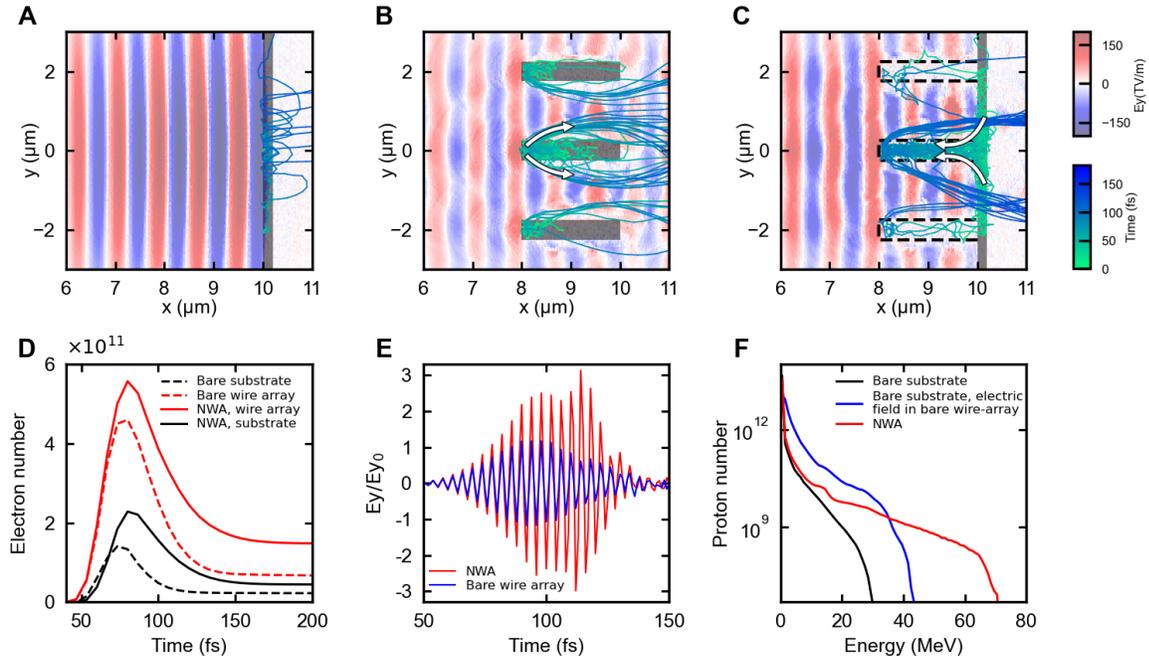

**Fig.3. Simulation results.** 2D slice of the transversal electric field $E_y$ at t = 69 fs and hot electron (> 10 MeV at t = 69 fs) trajectories in the x-y plane for three different cases of **(A)** bare substrate, **(B)** bare wire array, **(C)** combined NWA target. **(D)** Time evolution of the hot electron number with different target geometries. Solid lines (black: substrate, red: wire array) for the combined NWA target and dashed lines (black: bare substrate, red: bare wire array) for individual components. **(E)** Time evolution of the transversal electric field $E_y$ in the wire interval at x = 9 μm. **(F)** The proton spectra for three different cases of the bare substrate, the bare substrate with an introduced electric field, and the combined NWA target.

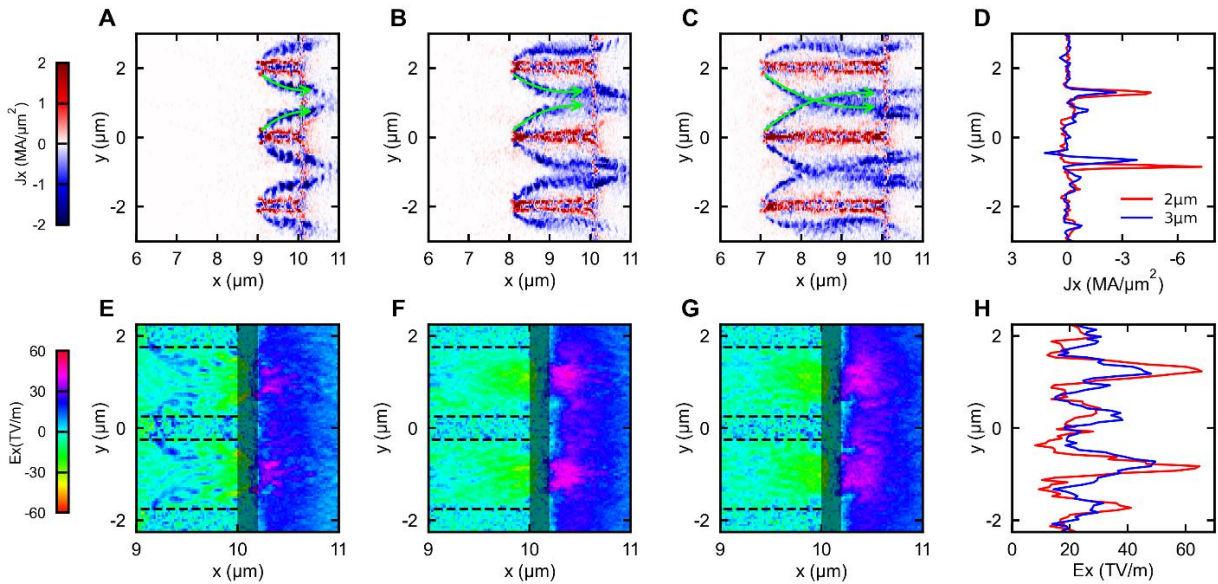

**Fig.4. The effect of the wire height on the longitudinal current and electric fields.**
Spatial distribution of the longitudinal current $J_x$ in the x-y plane at t = 69 fs for three wire heights of **(A)** 1 μm, **(B)** 2 μm and **(C)** 3 μm. The green arrows in (A)-(C) represent the electron bunch direction. **(D)** Transversal profiles of $J_x$ at x = 10.5 μm. Spatial distribution of the longitudinal electric field $E_x$ in the x-y plane at t = 69 fs for three wire heights of **(E)** 1 μm, **(F)** 2 μm and **(G)** 3 μm. **(H)** Transversal profiles of $E_x$ at x = 10.5 μm.

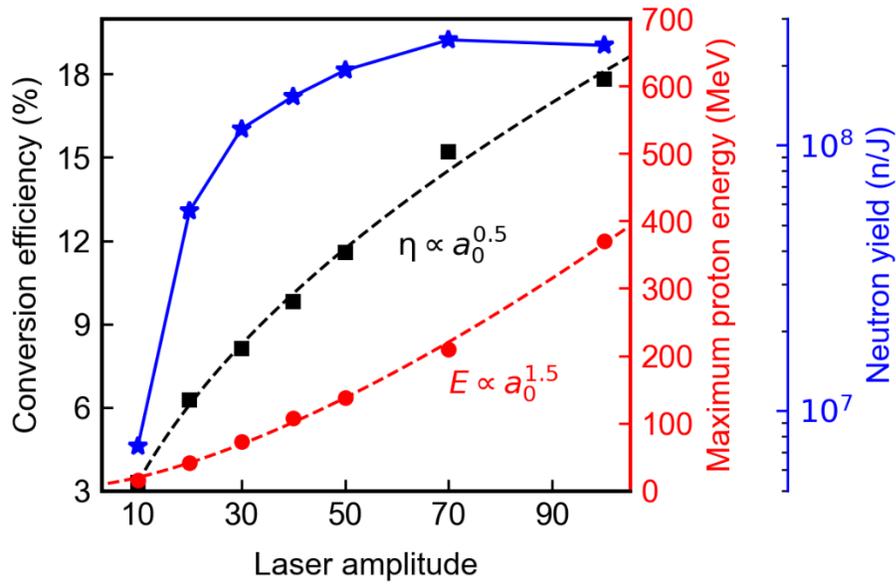

**Fig.5. Scaling with laser amplitude.** The proton conversion efficiency (black square), maximum proton energy (red circle) and neutron yield per joule (blue star) versus the input laser intensity for the 2-μm-height NWA target.

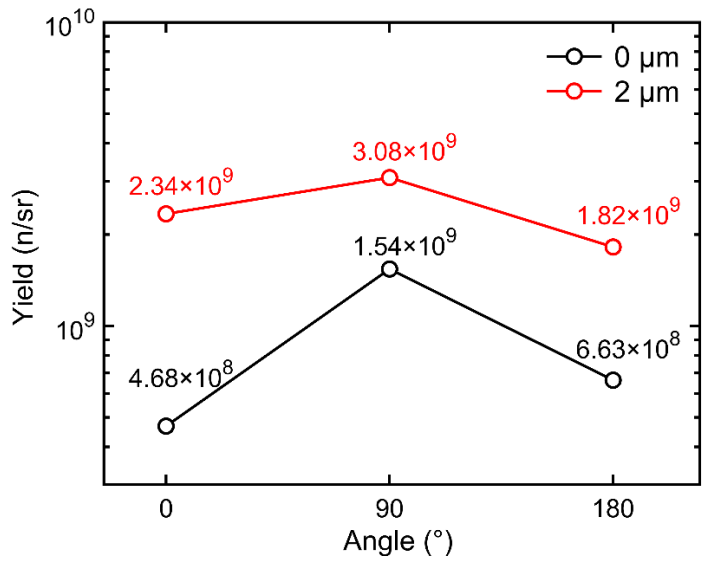

**Fig.6. Enhanced neutron yields characterized by BDs.** Neutron yields at three different angles of 0°, 90°, and 180° for two cases of the flat foil and the 2-μm-height NWA target.